\documentclass[12pt]{iopart}
\usepackage{graphicx}
\usepackage{comment}
\usepackage{subcaption}
\begin{document}

\title[Coherent Control of Evanescent Waves via Beam Shaping]{Coherent Control of Evanescent Waves via Beam Shaping}

\author{Nicholas J Savino$^{1}$\footnote{NJS, JML and WZ contributed equally to this work.}, Jacob M Leamer$^{1}$, Wenlei Zhang$^{1}$, Ravi K Saripalli$^{1,2}$, Ryan T Glasser$^{1}$ and Denys I Bondar$^{1}$}

\address{$^{1}$ Department of Physics and Engineering Physics, Tulane University, 6823 St. Charles Avenue, New Orleans, Louisiana 70118, USA \\ $^{2}$ Directed Energy Research Centre, Technology Innovation Institute, Abu Dhabi, UAE}
\eads{\mailto{rglasser@tulane.edu}, \mailto{dbondar@tulane.edu}}
\vspace{10pt}
\begin{indented}
\item[]October 2022
\end{indented}

\begin{abstract}
Evanescent waves are central to many technologies such as near-field imaging that beats the diffraction limit and plasmonic devices. Frustrated total internal reflection (FTIR) is an experimental method commonly used to study evanescent waves. In this paper, we shape the incident beam of the FTIR process with a Mach-Zehnder interferometer and measure light transmittance while varying the path length difference and interferometric visibility. Our results show that the transmittance varies with the path length difference and, thus, the intensity distribution of the shaped beam. Experiment and finite element method simulation produce results that agree. We also show, through simulations, that the transmittance can be controlled via other methods of beam shaping. Our work provides a proof-of-concept demonstration of the coherent control of the FTIR process, which could lead to advancements in numerous applications of evanescent waves and FTIR.
\end{abstract}

%
%
%
%
%
\section{Introduction}\label{sec:introduction}

Evanescent waves are inhomogeneous waves resulting from the continuity of the electric field under the conditions of total internal reflection (TIR). TIR occurs when a light wave propagates from an optically denser medium into one of lower density and the incidence angle is greater than the critical angle. Under such conditions, the electric field decays exponentially into the medium with lower optical density while the evanescent wave propagates parallel to the interface and, on average, no energy is carried across the interface. The penetration depth of the decaying field is on the order of magnitude of the wavelength of the incident light wave. The decaying field can be coupled into a third, optically dense medium to induce a forward-propagating wave by placing said medium after the TIR interface such that the distance between the two interfaces is close to the penetration depth. When this happens, energy may flow through the two interfaces into the third medium in what is known as frustrated total internal reflection (FTIR) \cite{hecht_optics_2015}. FTIR is commonly used in the study of evanescent waves.

Evanescent waves and FTIR have been known and studied since the formulation of Maxwell's equations more than 150 years ago. Traditionally, evanescent waves were considered to be elusive and merely a by-product of TIR. However, with technological advances in lasers, nano-scale fabrication, and photonics, evanescent waves and FTIR have been found useful in several different fields of study. The invention of the near-field scanning optical microscope \cite{pohl_optical_1984,betzig_breaking_1991,reddick_new_1989}, breaking the diffraction limit, utilized and bettered our understanding of FTIR. Other applications include near field sensing \cite{novotny_principles_2006}, non-radiative energy transfer through the F\"{o}rster process \cite{andrew_forster_2000}, and manipulation of particles using near-field photonic forces \cite{richards_near-field_2004}. Recent studies in near-field optics have shown that evanescent waves can carry transverse spin angular momentum \cite{marrucci_spin_2015,bliokh_spinorbit_2015,aiello_transverse_2015}, leading to the photonic quantum spin Hall effect \cite{bliokh_extraordinary_2014,bliokh_quantum_2015,van_mechelen_universal_2016}. Evanescent waves of vector beams and non-paraxial beams have also been subjects of interest due to their potential applications in photonics and sensing \cite{shchegrov_far-field_1999,frederique_de_fornel_evanescent_2001,helseth_focusing_2010,chen_evanescent_2013}. More notably, FTIR is considered to be a close analogy to quantum tunneling \cite{doi:10.1119/1.14514,PhysRevE.48.632,1975ApPhy...6..131H}, the further study of which may give us insight into both classical optics and quantum mechanics.

In this paper, we shape the incident beam with the interference pattern from a Mach-Zehnder interferometer (MZI). We measure the transmittance of the induced forward-propagating wave while varying the path length difference between the two arms of the MZI and the interferometric visibility. We find that the transmittance can be enhanced or suppressed depending on the phase difference between the waves in the incident interference pattern. We show, through simulation, that this control of the transmittance comes from beam shaping of the incident beam. Our results serve as a proof-of-principle demonstration for the coherent control of FTIR. Enhancement of the FTIR transmittance could improve the sensitivity of spectroscopy and imaging techniques based on evanescent waves and FTIR. More accurate control of the FTIR transmittance could also lead to advancements in plasmonics technology based on the photonic quantum spin Hall effect \cite{mittal_photonic_2019,dai_plasmonic_2019,wan_controlling_2020,bahari_photonic_2021,liu_enhanced_2021}. Previously, controlling and shaping the evanescent field by metamaterials has been proposed and demonstrated \cite{ginis_remote_2020}.

\section{Experimental measurement}

\begin{figure}[htbp]
    \centering
    \includegraphics[width=.8\textwidth]{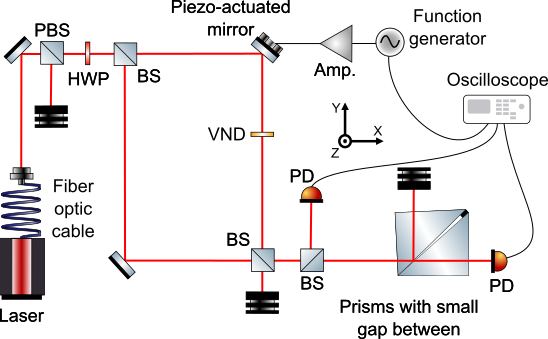}
    \caption{\small{Diagram of the experimental setup. A near-infared laser ($\lambda \approx 795$nm) is used to generate the light, which is collimated by a single mode optical fiber. The light is passed through a Mach-Zehnder interferometer, creating an interference pattern which is then passed through the FTIR prism setup. Some light is split off before the prism setup in order to determine the incident light intensity. Abbreviations: PBS = polarizing beam splitter, HWP = half-waveplate, VND = variable neutral density filter, PD = photodetector, Amp. = amplifier. The $x$ (right), $y$ (top of the page), and $z$ (out of the page) directions are indicated.}}
    \label{fig:experimental setup}
\end{figure}

Figure~\ref{fig:experimental setup} shows our experimental setup for FTIR with an interference pattern from a MZI as incident light. Two large, glass, uncoated prisms are clamped together with a small aluminum foil shim, allowing for pressure to be concentrated in a small area and the separation between the prisms to be on the order of tens of micrometers \cite{voros_simple_2008}. The incidence angle into the hypotenuse of the first prism is greater than the critical angle ($\sim 42^\circ$) and TIR occurs between the first prism and the air gap. The exponentially decaying field of the evanescent wave is coupled into the second prism to produce a forward-propagating wave via FTIR. The optical path lengths from the final beamsplitter to the two photodetectors are made to be equal. We verify that we are in fact observing FTIR by translating the incidence beam in the $+y$-direction, which corresponds to a linear increase in the air gap width. This results in an exponential decay of the total intensity on the output side of the FTIR prisms as the air gap width increases, which confirms FTIR.

The incidence light into the FTIR prisms is an interference pattern between two overlapping Gaussian beams in a MZI. A piezoelectric steering mirror is placed in one arm to change the path length difference between the two arms. The interferometric visibility, defined as

\begin{equation}
v = \frac{I_\mathrm{max}-I_\mathrm{min}}{I_\mathrm{max}+I_\mathrm{min}},
\end{equation}
where $I_\mathrm{max}$ and $I_\mathrm{min}$ are the respective maximum and minimum intensity of the output of the MZI as the path length difference changes from 0 to $\lambda$, can be varied by attenuating one arm of the MZI with a variable neutral density filter. We are able to obtain a maximum interferometric visibility of $v \approx 70\%$ in our experiment. We measured the transmittance of energy of the FTIR process, defined as the ratio of the total transmitted intensity to the total incident intensity of the FTIR prisms, of both $y$-polarized and $z$-polarized incidence light.

\begin{figure}[htbp]
    \centering
    \includegraphics[width=\linewidth]{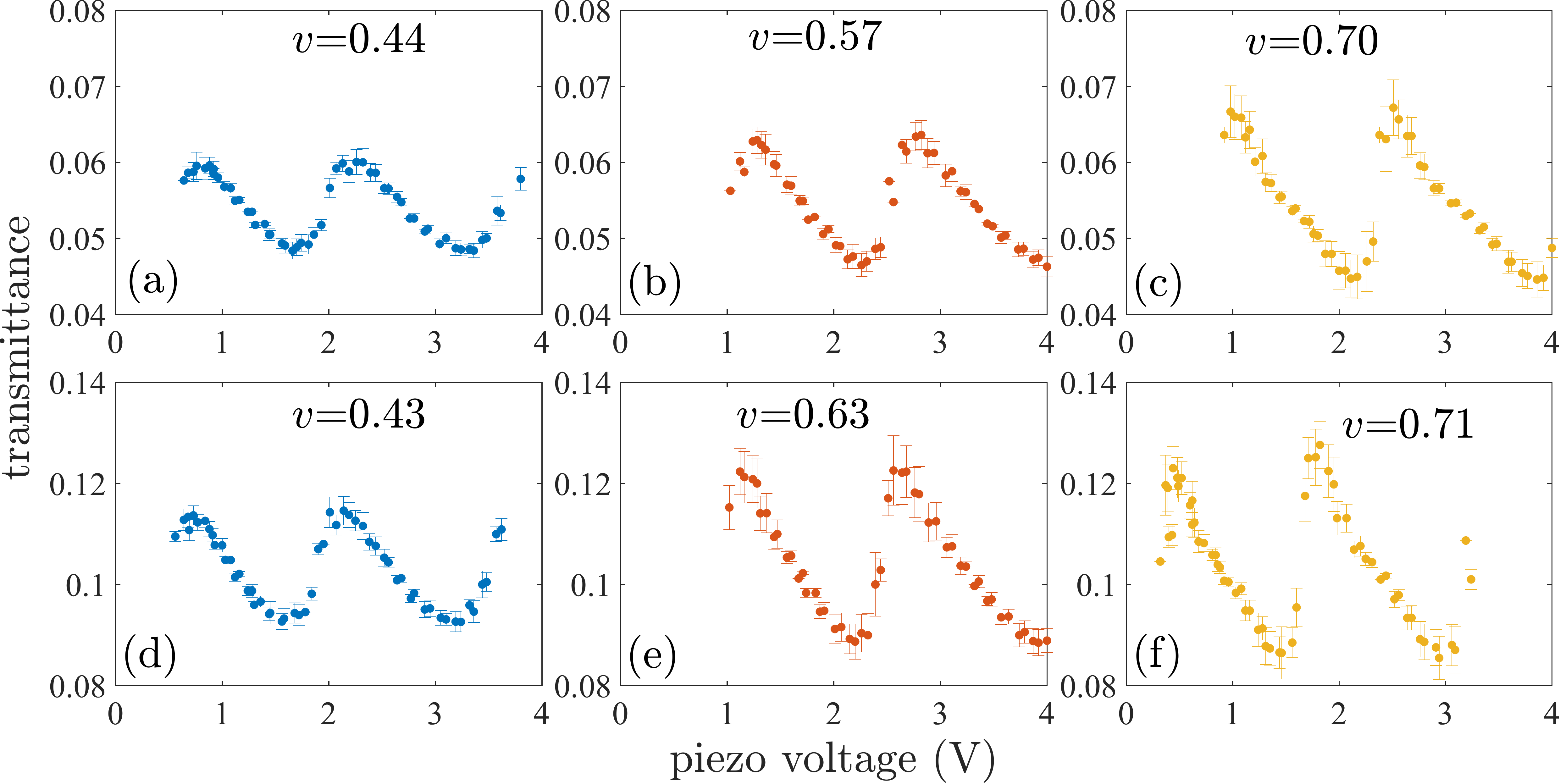}
    \caption{\small{Measured transmittances of the evanescent waves generated via FTIR for (a), (b), (c) $z$-polarized light and (d), (e), (f) $y$-polarized light with different visibility levels of the incident interference pattern. The error bars represent 95\% confidence interval of repeated measurements.}}
    \label{fig:results}
\end{figure}

We experimentally measure the transmittance of the FTIR process with the interference pattern from a MZI as the incident beam using the setup shown in figure~\ref{fig:experimental setup}. The transmittance is the ratio of the total intensity of the forward-propagating wave to the total incident intensity. We measure the transmittance while varying the voltage on the piezoelectric driver for different incident polarizations and interferometric visibility. 
In the experimental setup, the varying piezo voltage leads to varying path-length difference in the two arms of the MZI, which ultimately results in different intensity profiles of the interference pattern incident on the FTIR prisms. Figure~\ref{fig:results} shows the experimentally measured transmittance. These measurements show it is possible to change the transmittance by varying the phase difference between the two beams that make up the incident interference pattern. Another point of interest to note is that $y$-polarized incident light has a higher transmittance for the same phase difference and visibility than $z$-polarized incident light. We notice that there appears to be a small shift in the x-axis between different trials. This is not due to a phase sift, but rather due to the nature of our method of varying phase with a piezo motor. The zero point of the piezo does not necessarily reset to the same phase for every trial, producing this artifact. We compare these measurements to the simulation results obtained in the next section.

\section{Simulation\label{sec:simulation}}

\begin{figure}[htbp]
    \centering
    \includegraphics[width=.85\linewidth]{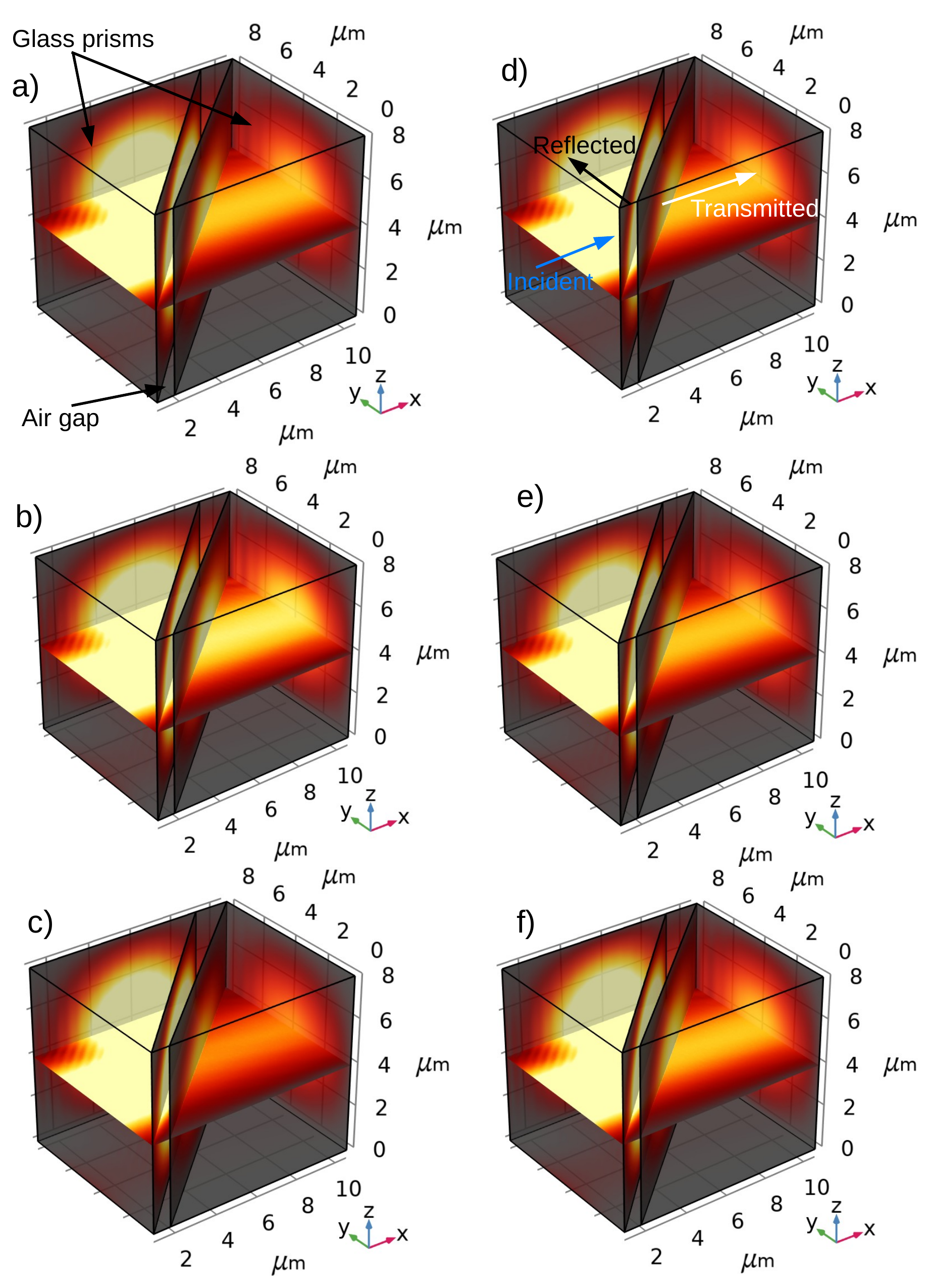}
    \caption{\small{Surface plots of the electric field norm through a system of two prisms separated by an air gap for two incident beams that are $y$-polarized and offset by 0.04 \textmu m in the (a) $y$-direction with no phase difference, (b) $y$-direction with a phase difference of 175$^\circ$, (c) $y$-direction with a phase difference of 185$^\circ$, (d) $z$-direction with no phase difference, (e) $z$-direction with a phase difference of 175$^\circ$, and (f) $z$-direction with a phase difference of 185$^\circ$.  The results here are normalized.  These plots demonstrate that the forward propagating wave depends on the phase difference between the two beams only when they are offset in the $y$-direction, not when they are offset in the $z$-direction.}}
    \label{fig:simulation}
\end{figure}

We use the COMSOL Wave Optics \cite{COMSOL} software to simulate the measurements done with the experimental setup shown in figure~\ref{fig:experimental setup}. In the simulation, the light propagates in the $x$-direction through two glass prisms that have refractive index $n=1.5$ and that are separated by an air gap as displayed in figure~\ref{fig:simulation} (a).  We chose to use a wavelength $\lambda = 795$ nm and set the prisms to measure 8 \textmu m $\times$ 8 \textmu m $\times$ 8 \textmu m so that the simulations could be run in a reasonable amount of time.  The beams were set to have a waist size $w_0 = 2$ \textmu m to reduce the presence of boundary effects in our results.  The prisms were also oriented so that the thickness of the air gap was one $\lambda$ on one side of the system and $2\lambda$ on the other.  For all of our simulations, the hypotenuse of the first prism is set to make a 45$^\circ$ angle with the $x$-axis so that the incident angle is greater than the critical angle ($\sim 42^\circ$) of the FTIR process.
\par
The surface plots of the norm of the electric field in figure~\ref{fig:simulation} are generated by simulating the propagation of two Gaussian beams with offset optical axes and a phase difference between them through our two prism setup.  The offset was chosen to be small, but non-zero to reflect the fact that the beams cannot be perfectly aligned in the experiment.  In figure~\ref{fig:simulation} (a)-(c), the beams are offset in the $y$-direction by 0.04 \textmu m, while in figure~\ref{fig:simulation} (d)-(f), they are offset in the $z$-direction by 0.04 \textmu m.  In figure~\ref{fig:simulation}, the phase difference between the two incident beams is zero for (a) and (d), $175^\circ$ for (b) and (e), and $185^\circ$ for (c) and (f). The plots in figure~\ref{fig:simulation} show that the phase difference between the two incident beams affect the forward-propagating wave only when they are offset in the $y$-direction, not when they are offset in the $z$-direction. The intuition behind this phenomenon is that the two incident beam travel through different optical path lengths in the first prism before arriving at the air gap when they are offset in the $y$-direction, but they see the same optical path length when they are offset in the $z$-direction.

\par
\begin{figure}[htbp]
    \centering
    \begin{subfigure}[b]{.45\linewidth}
        \centering
        \includegraphics[width=\linewidth]{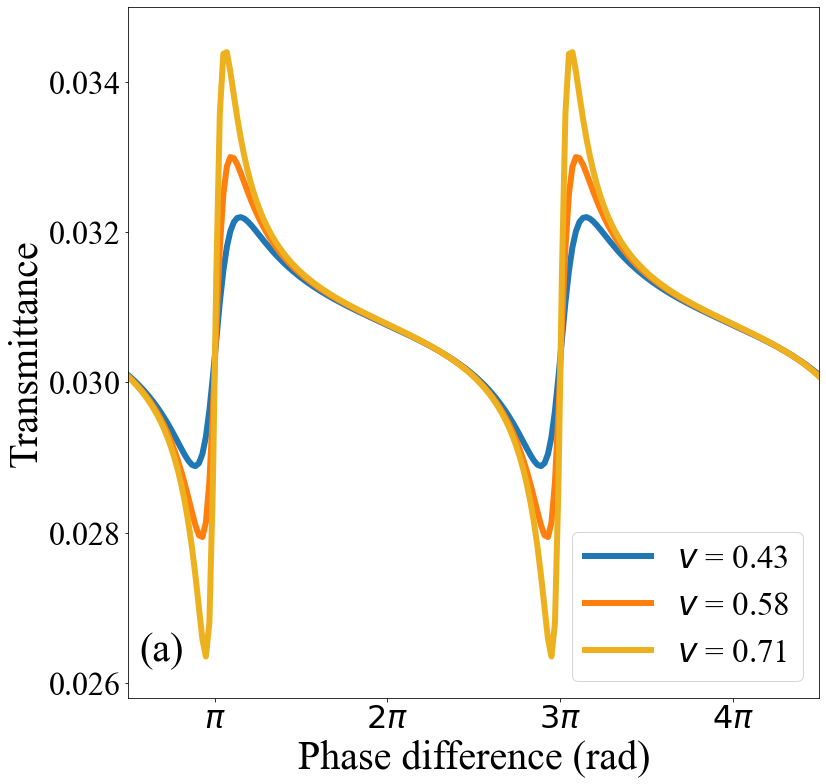}
    \end{subfigure}
    \begin{subfigure}[b]{.45\linewidth}
        \centering
        \includegraphics[width=\linewidth]{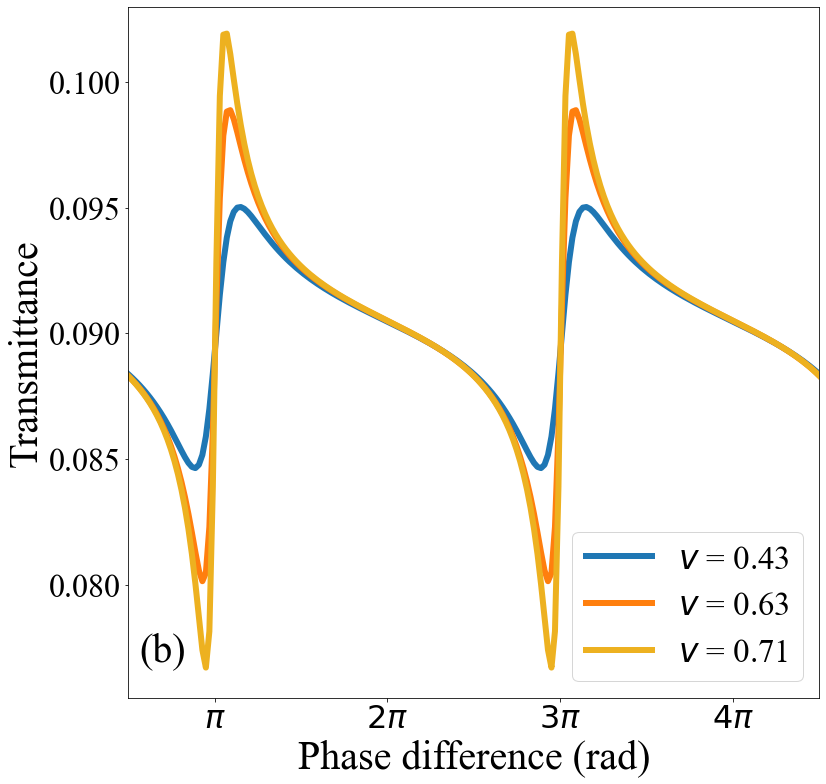}
    \end{subfigure}
    \caption{\small{Simulated transmittance of energy via FTIR with varying gap width for (a) $z$-polarized and (b) $y$-polarized light and different visibility values of the incident shaped beam.}}
    \label{fig:simulated}
\end{figure}

We then want to investigate the dependence of transmittance on all values of the phase difference $\phi$ and the interferometric visibility $v$. Since FTIR is a linear optics phenomena, we can do this by first simulating the propagation of a single Gaussian beam $\vec{E}_1$ and a single shifted beam $\vec{E}_2$ separately and then add the results together to obtain the total field
\begin{equation}
    \vec{E}_T = \vec{E}_1 + ae^{i\phi}\vec{E}_2,
\end{equation}
where $\phi$ is the phase difference between the beams and $a$ is a coefficient that allows us to tune the amplitude of $\vec{E}_2$, leading to a change in $v$.  For these simulations, the offset between $\vec{E}_1$ and $\vec{E}_2$ is 0.04 \textmu m in the $y$-direction.

This approach is used to obtain results when $\vec{E}_1$ and $\vec{E}_2$ are $y$-polarized and $z$-polarized. Figure~\ref{fig:simulated} shows the simulated results. Here, the transmittance is the ratio of the total intensity of the forward-propagating wave to the total intensity of the incident waves. The simulation results shown in figure~\ref{fig:simulated} has a qualitative agreement with the experimental measurements shown in figure~\ref{fig:results}. In both the simulation and experiment, $y$-polarized incident light has a higher transmittance for the same phase difference and visibility. The difference between the simulation and experiment is mainly due to the different wavelength and geometry of the prisms used in the simulation to speed up calculations. 
Both experimental measurement and simulation result show that it is possible to modulate the transmittance in FTIR by varying the phase difference and higher visibility increases the range of this modulation.

\par

\begin{table}[htbp]
\caption{\small{Transmittance obtained from simulating the propagation of a single Hermite-Gauss beam through the FTIR prisms with different mode indices.}}
    \centering
    \begin{tabular}{cccc}
        \hline
        $HG_{n_y,n_z}$ & $n_z=0$ & $n_z=1$ & $n_z=2$ \\
        \hline
        $n_y=0$ & 0.090 & 0.084 & 0.076 \\
        $n_y=1$ & 0.203 & 0.193 & 0.177 \\
        $n_y=2$ & 0.303 & 0.292 & 0.273 \\
        \hline
    \end{tabular}

    \label{tab:hg_mode}
\end{table}

The previous simulations suggest that the transmittance in the FTIR process could be controlled by beam shaping of the incident beam. To confirm that other methods of beam shaping can be used to control this process, we simulate the propagation of single Hermite-Gauss beams through our two-prism setup.  The Hermite-Gauss modes were chosen because the nodes that occur in the beam profile as one moves to higher order modes resemble the offsets between beams that were used in our previous simulations. Table~\ref{tab:hg_mode} shows the transmittance obtained by simulating the propagation of a single Hermite-Gauss beam through the FTIR prisms for mode indices ranging from 0 to 2.  These results show that for any given $z$-mode index, the transmittance is larger for a larger $y$-mode index.  In contrast, it is also clear that for any given $y$-mode index, the transmittance is smaller for a larger $z$-mode index.  This confirms that control of evanescence is more generally a beam shaping phenomenon.

\begin{figure}[htbp]
    \centering
    \includegraphics[width=0.8\linewidth]{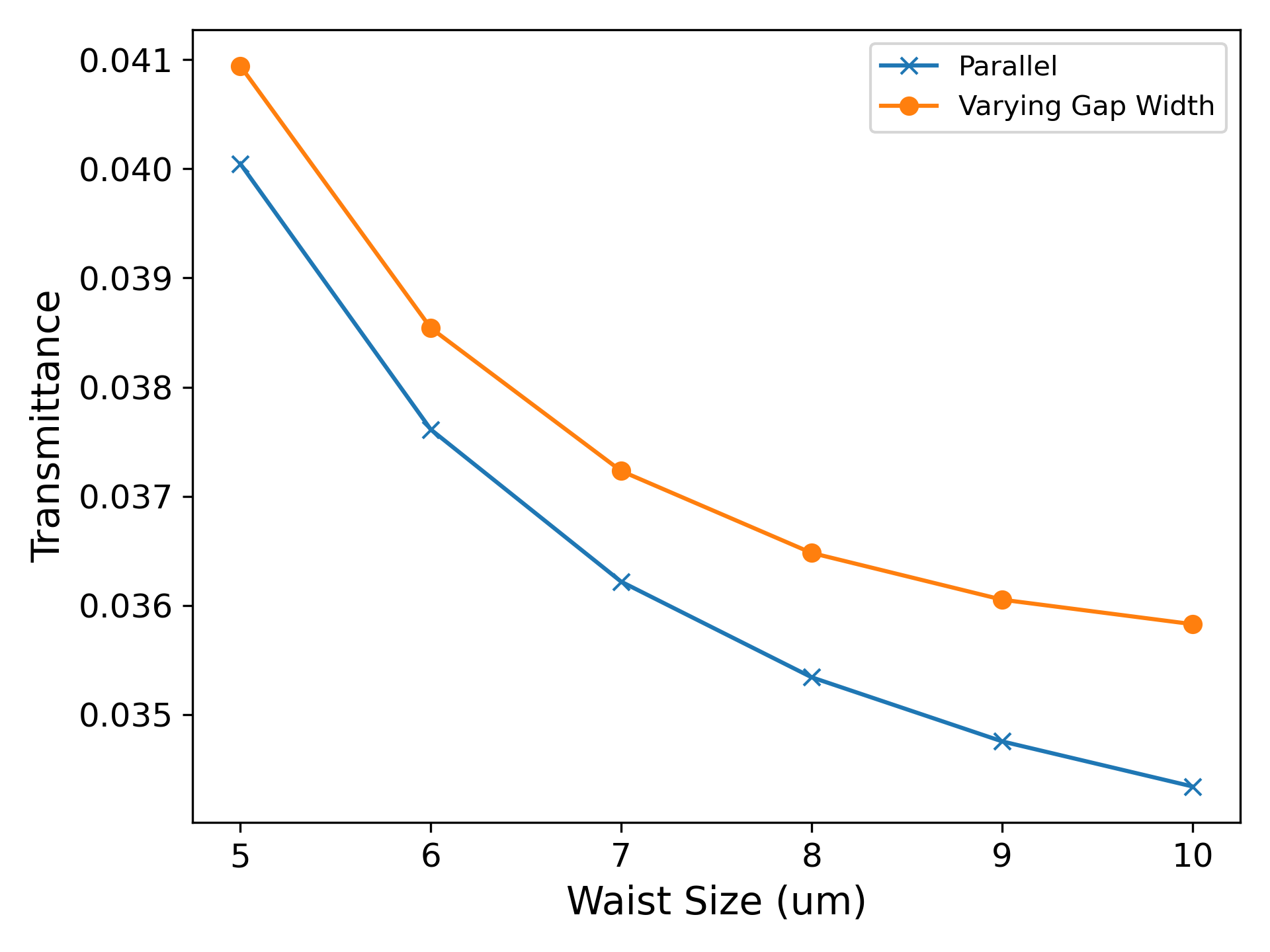}
    \caption{\small{Transmittance of a single Gaussian beam through the FTIR prisms for different waist sizes when the the prisms are oriented so that their hypotenuses are parallel (blue crosses) and when the prisms are oriented so that the thickness of the gap varies from $\lambda$ to $2\lambda$ (orange dots).  These results show that even simple methods of beam shaping can be used to control the strength of evanescence and that this is robust to the orientation of the prisms.}}
    \label{fig:waist_size}
\end{figure}

As another confirmation that control of evanescence is achieved by beam shaping, we consider the simplest case of beam shaping - changing the waist size of a Gaussian beam. We then simulate the propagation of a single Gaussian beam with different waist sizes through the FTIR prisms.  In this simulation, the wavelength is $\lambda = 795$ nm and the size of the prisms are changed to 60 \textmu m. We consider two different orientations of the prisms. In the first case, the prisms are oriented so that their hypotenuses are parallel and separated by an air gap that is 1.5$\lambda$ thick.  In the second case, they are oriented so that the thickness of the air gap varies from $\lambda$ on one end to 2$\lambda$ on the other, same as that in the simulations in figure~\ref{fig:simulation}. The second case is closer to our experimental setup shown in figure~\ref{fig:experimental setup}. We would like to see if such a varying gap width has an effect on the FTIR process by comparing these two cases. Figure~\ref{fig:waist_size} shows the transmittance of the Gaussian beam with different waist sizes for both cases. In figure~\ref{fig:waist_size}, the transmittance decreases monotonically as the waist size increases and this general trend is the same for both orientation of the prisms. These results suggest that the varying gap width has no major effect on the behavior of the transmittance.

\begin{figure}[htbp]
    \centering
    \includegraphics[width=0.8\linewidth]{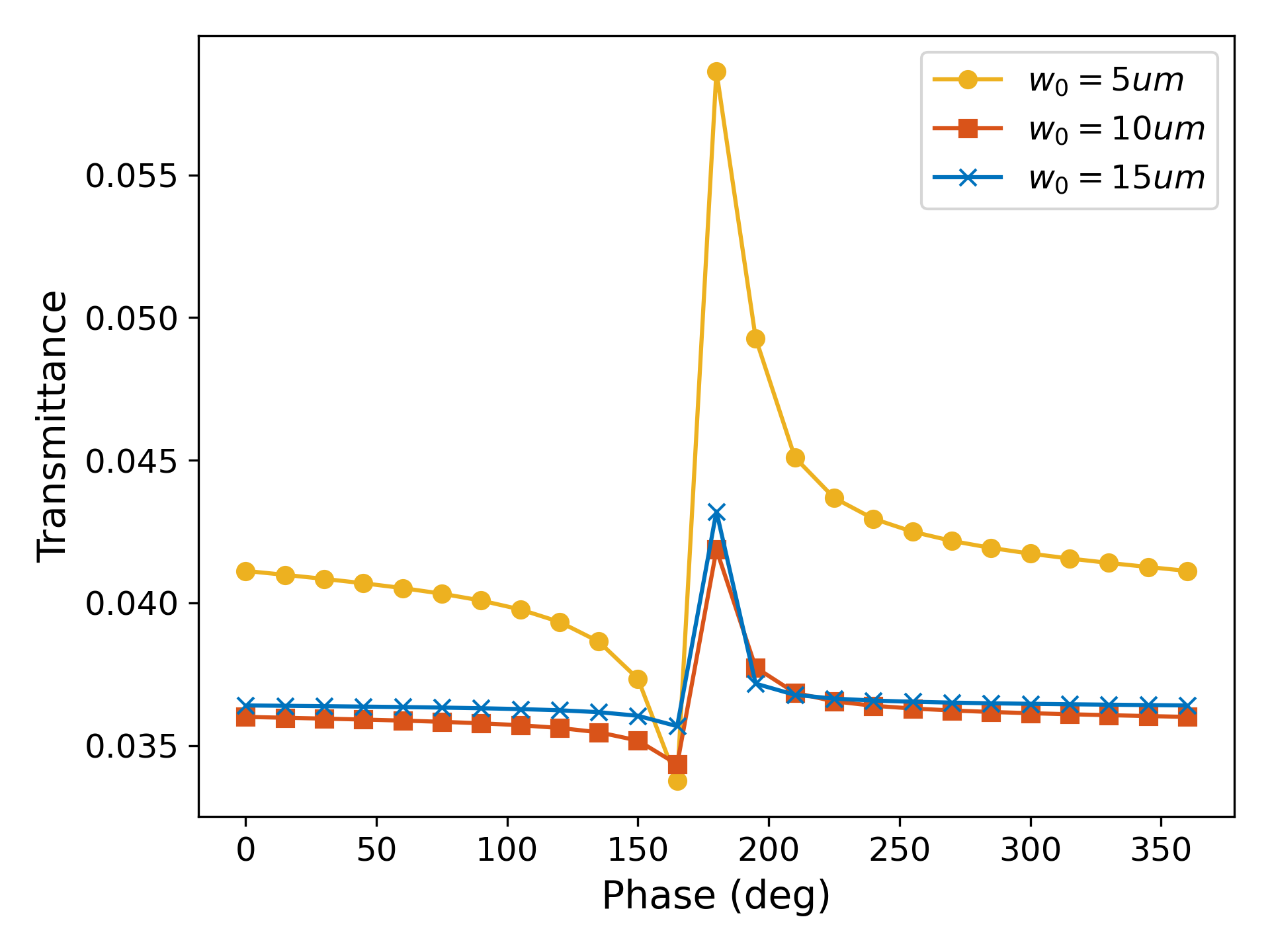}
    \caption{\small{Transmittance of two offset Gaussian beams through the FTIR prisms vs the phase difference between the beams for different waist sizes $w_0$ of the beams.  Both beams always have the same value for $w_0$.  This demonstrates that controlling evanescence by changing the beam shape depends on the waist size of the beam.}}
    \label{fig:comp_waist_size}
\end{figure}

To develop a better intuition for why the transmittance of evanescent waves can be controlled by changing the beam shape, we consider the form of a Gaussian beam propagating in the $x$-direction and polarized in the $y$-direction
\begin{equation}
    \vec{E}(x, r) = E_0\frac{w_0}{w(x)}e^{-\frac{r^2}{w(x)^2}}e^{-ikx}e^{-ik\frac{r^2}{2R(x)}}e^{i\psi(x)}\hat{y},
\end{equation}
where $E_0$ is the initial amplitude, $w_0$ is the waist size of the beam, $w(x)$ is the radius of the beam at x, $r$ is the radial distance from the optical axis, $k$ is the wavenumber, $R(x)$ is the radius of curvature of the wavefront, and $\psi(x)$ is the Guoy phase.  The Rayleigh range
\begin{equation}
    x_R = \frac{\pi w_0^2 n}{\lambda},
\end{equation}
relates $w(x)$, $R(x)$, and $\psi(x)$ to $w_0$ in the following ways:
\begin{equation}
    w(x) = w_0\sqrt{1 + \left(\frac{x}{x_R}\right)^2},
\end{equation}
\begin{equation}
    R(x) = x\left(1+\left(\frac{x_R}{x}\right)^2\right),
\end{equation}
\begin{equation}
    \psi(x) = \arctan{\frac{x}{x_R}}.
\end{equation}
Now if $w_0 \rightarrow \infty$, we see that $w(x) \rightarrow \infty$, $R(x) \rightarrow \infty$, $\psi(x) \rightarrow 0$, and $\frac{w_0}{w(x)} \rightarrow 1$.  This leaves us with a regular plane wave, which doesn't allow for controlling the transmittance of evanescent waves.  Thus, it appears that controlling evanescence by changing the beam shape depends on the waist size of the beam.  We tested this by simulating the propagation of two Gaussian beams through the FTIR prisms in 2D as the phase difference between the beams is changed.  We repeat this for different values of $w_0$, though in each case both beams have the same $w_0$.  In the simulation, $\lambda = 795$ nm and the beams are offset by 0.2 \textmu m.  The prisms are 60 \textmu m x 60 \textmu m and are oriented so that the air gap has thickness $\lambda$ on one edge and thickness 2$\lambda$ on the other.  The transmittance of the beams is plotted against the phase difference between them for the cases $w_0 = 5$ \textmu m, $w_0 = 10$ \textmu m, and $w_0 = 15$ \textmu m in figure \ref{fig:comp_waist_size}.  The curve for the $w_0 = 5$ \textmu m case is very similar to the results in figure \ref{fig:simulated}, but the curve flattens out as $w_0$ is increased.  This confirms that the ability to modulate evanescence by changing the shape of the beam depends on the beam's waist size.  In particular, a smaller $w_0$ allows for greater modulation.

\section{Discussion}

We measured the transmittance of energy in an FTIR process with an interference pattern as incident light. Both experimental measurement and simulation result show that the transmittance can be controlled depending on the intensity profile of the incident beam. We also show, through simulation, that the transmittance can be controlled by other means of beam shaping. Our results suggest that more sophisticated designs of coherent control of the FTIR process can be devised. This effect, to the best of our knowledge, have not been previously observed and shows promise for applications where FTIR is utilized, such as near-field sensing \cite{novotny_principles_2006} and non-radiative energy transfer \cite{andrew_forster_2000}.

\section*{Funding}
U. S. Office of Naval Research (N000141912374); Defense Advanced Research Projects Agency (D19AP00043); U.S. Army Research Office (W911NF-19-1-0377).

\section*{Acknowledgments}
This material is based upon research supported by, or in part by, the U. S. Office of Naval Research under award number N000141912374. This work was also supported by the Defense Advanced Research Projects Agency (DARPA) grant number D19AP00043 under mentorship of Dr. Joseph Altepeter. D.I.B. is also supported by the U.S. Army Research Office (ARO) under grant W911NF-19-1-0377. The views and conclusions contained in this document are those of the authors and should not be interpreted as representing the official policies, either expressed or implied, of DARPA, ONR, ARO, or the U.S. Government. The U.S. Government is authorized to reproduce and distribute reprints for Government purposes notwithstanding any copyright notation herein.

\section*{Disclosures}
The authors declare no conflicts of interest.

\section*{Data availability} Data underlying the results presented in this paper are not publicly available at this time but may be obtained from the authors upon reasonable request.

\section*{References}
\bibliography{literature}

\providecommand{\newblock}{}
\begin{thebibliography}{10}
\expandafter\ifx\csname url\endcsname\relax
  \def\url#1{{\tt #1}}\fi
\expandafter\ifx\csname urlprefix\endcsname\relax\def\urlprefix{URL }\fi
\providecommand{\eprint}[2][]{\url{#2}}

\bibitem{hecht_optics_2015}
Hecht E 2015 {\em Optics\/} 5th ed (Pearson) ISBN 978-0-13-397722-6

\bibitem{pohl_optical_1984}
Pohl D~W, Denk W and Lanz M 1984 {\em Applied Physics Letters\/} {\bf 44}
  651--653 ISSN 0003-6951, 1077-3118

\bibitem{betzig_breaking_1991}
Betzig E, Trautman J~K, Harris T~D, Weiner J~S and Kostelak R~L 1991 {\em
  Science\/} {\bf 251} 1468--1470 ISSN 0036-8075, 1095-9203

\bibitem{reddick_new_1989}
Reddick R~C, Warmack R~J and Ferrell T~L 1989 {\em Physical Review B\/} {\bf
  39} 767--770 ISSN 0163-1829

\bibitem{novotny_principles_2006}
Novotny L and Hecht B 2006 {\em Principles of {Nano}-{Optics}\/} (Cambridge:
  Cambridge University Press)

\bibitem{andrew_forster_2000}
Andrew P and Barnes W~L 2000 {\em Science\/} {\bf 290} 785--788

\bibitem{richards_near-field_2004}
Richards D, Zayats A, Nieto-Vesperinas M, Chaumet P~C and Rahmani A 2004 {\em
  Philosophical Transactions of the Royal Society of London. Series A:
  Mathematical, Physical and Engineering Sciences\/} {\bf 362} 719--737

\bibitem{marrucci_spin_2015}
Marrucci L 2015 {\em Nature Physics\/} {\bf 11} 9--10 ISSN 1745-2473, 1745-2481

\bibitem{bliokh_spinorbit_2015}
Bliokh K~Y, Rodríguez-Fortuño F~J, Nori F and Zayats A~V 2015 {\em Nature
  Photonics\/} {\bf 9} 796--808 ISSN 1749-4893

\bibitem{aiello_transverse_2015}
Aiello A, Banzer P, Neugebauer M and Leuchs G 2015 {\em Nature Photonics\/}
  {\bf 9} 789--795 ISSN 1749-4885, 1749-4893

\bibitem{bliokh_extraordinary_2014}
Bliokh K~Y, Bekshaev A~Y and Nori F 2014 {\em Nature Communications\/} {\bf 5}
  3300 ISSN 2041-1723

\bibitem{bliokh_quantum_2015}
Bliokh K~Y, Smirnova D and Nori F 2015 {\em Science\/} {\bf 348} 1448--1451
  ISSN 0036-8075, 1095-9203

\bibitem{van_mechelen_universal_2016}
Van~Mechelen T and Jacob Z 2016 {\em Optica\/} {\bf 3} 118 ISSN 2334-2536

\bibitem{shchegrov_far-field_1999}
Shchegrov A~V and Carney P~S 1999 {\em JOSA A\/} {\bf 16} 2583--2584 ISSN
  1520-8532

\bibitem{frederique_de_fornel_evanescent_2001}
{Frederique de Fornel} 2001 {\em Evanescent {Waves}: {From} {Newtonian}
  {Optics} to {Atomic} {Optics}\/} 1st ed Springer {Series} in {Optical}
  {Sciences} (Heidelberg: Springer Berlin) ISBN 978-3-642-08513-0

\bibitem{helseth_focusing_2010}
Helseth L 2010 {\em Optics Communications\/} {\bf 283} 29--33 ISSN 00304018

\bibitem{chen_evanescent_2013}
Chen R~P and Li G 2013 {\em Optics Express\/} {\bf 21} 22246 ISSN 1094-4087

\bibitem{doi:10.1119/1.14514}
Zhu S, Yu A~W, Hawley D and Roy R 1986 {\em American Journal of Physics\/} {\bf
  54} 601--607 (\textit{Preprint} \eprint{https://doi.org/10.1119/1.14514})
  \urlprefix\url{https://doi.org/10.1119/1.14514}

\bibitem{PhysRevE.48.632}
Enders A and Nimtz G 1993 {\em Phys. Rev. E\/} {\bf 48}(1) 632--634
  \urlprefix\url{https://link.aps.org/doi/10.1103/PhysRevE.48.632}

\bibitem{1975ApPhy...6..131H}
{Hupert} J~J 1975 {\em Applied Physics\/} {\bf 6} 131--149

\bibitem{mittal_photonic_2019}
Mittal S, Orre V~V, Leykam D, Chong Y and Hafezi M 2019 {\em Physical Review
  Letters\/} {\bf 123} 043201

\bibitem{dai_plasmonic_2019}
Dai Y and Petek H 2019 {\em ACS Photonics\/} {\bf 6} 2005--2013

\bibitem{wan_controlling_2020}
Wan R~G and Zubairy M~S 2020 {\em Physical Review A\/} {\bf 101} 033837

\bibitem{bahari_photonic_2021}
Bahari B, Hsu L, Pan S~H, Preece D, Ndao A, El~Amili A, Fainman Y and Kanté B
  2021 {\em Nature Physics\/} {\bf 17} 700--703 ISSN 1745-2481

\bibitem{liu_enhanced_2021}
Liu J, Liu J, Li X, Li X, Tao J, Tao J, Dong D, Dong D, Liu Y, Liu Y, Liu Y, Fu
  Y and Fu Y 2021 {\em Optics Letters\/} {\bf 46} 2537--2540 ISSN 1539-4794

\bibitem{ginis_remote_2020}
Ginis V, Piccardo M, Tamagnone M, Lu J, Qiu M, Kheifets S and Capasso F 2020
  {\em Science\/} {\bf 369} 436--440 ISSN 0036-8075, 1095-9203

\bibitem{voros_simple_2008}
Vörös Z and Johnsen R 2008 {\em American Journal of Physics\/} {\bf 76}
  746--749 ISSN 0002-9505, 1943-2909

\bibitem{COMSOL}
{COMSOL} {Multiphysics} \textregistered v. 6.0. {COMSOL} AB, Stockholm, Sweden
  \urlprefix\url{www.comsol.com}

\end{thebibliography}

\end{document}